\documentclass[twocolumn]{aastex701}

\usepackage{natbib,xspace}

\usepackage{multirow}

\usepackage{adjustbox}
\usepackage{booktabs}
\usepackage{tabularx}
\usepackage{graphicx}
\usepackage{bm}

\usepackage{soul}

\graphicspath{{/Users/laurynwilliams/Desktop/UW/research/writing/MANGA/submission/}}

\newcommand{\dtzos}{{dTŻOs}\xspace}
\newcommand{\dtzo}{{dTŻO}\xspace}
\newcommand{\manga}{\texttt{MANGA}\xspace}
\newcommand{\changa}{\texttt{ChaNGa}\xspace}
\newcommand{\tzos}{TŻOs\xspace}
\newcommand{\mesa}{\texttt{MESA}\xspace}

\shortauthors{Williams et al.}

\begin{document}

\title{Neutron Star-Main Sequence Collisions Robustly Form Dynamically Stable Thorne-Żytkow Objects}

\correspondingauthor{Lauryn E. Williams}

\author[orcid=0000-0002-0604-5440,gname='Lauryn E.',sname='Williams']{Lauryn E. Williams}
\affiliation{Department of Astronomy, University of Washington, Seattle, WA 98195, USA}
\email[show]{laurynwi@uw.edu}

\author[0000-0002-2137-2837]{Philip Chang}
\affiliation{Department of Physics, University of Wisconsin, Milwaukee, WI 53211, USA}
\email{chang65@uwm.edu}

\author[0000-0003-2184-1581]{Emily M. Levesque}
\affiliation{Department of Astronomy, University of Washington, Seattle, WA 98195, USA}
\email{emsque@uw.edu}

\author[0000-0001-5510-2803]{Thomas R. Quinn}
\affiliation{Department of Astronomy, University of Washington, Seattle, WA 98195, USA}
\email{trq@uw.edu}

\begin{abstract} 

Thorne-Żytkow Objects (\tzos) are hypothetical hybrid stars with a neutron star at the core of a large, diffuse envelope. \tzos may be formed when a newly formed neutron star that is kicked by its supernova collides with its main-sequence companion. Using a moving-mesh (MM) hydrodynamics solver integrated into the parallel-code Charm N-body GrAvity solver, we demonstrate that these ``impact scenario'' formation processes robustly form \tzos for periastron distances less than one stellar radius. These \tzos are dynamically stable and they can serve as initial models for further evolutionary studies. 
\end{abstract}


\section{Introduction} \label{sec:intro}

Thorne-Żytkow Objects (\tzos) are proposed as a hypothetical class of stars with a neutron star (NS) at the core of a large, diffuse envelope \citep{thorne_red_1975,thorne_stars_1977}. Massive \tzos are supported by the interrupted-rapid-proton process \citep{biehle_highmass_1991a,cannon_massive_1993}, a process that is uniquely possible thanks to the extremely high temperatures and fully convective envelopes present in these stars' interiors. This leads to heavy elements such as rubidium and molybdenum that are significantly enhanced over typical stellar abundances \citep{biehle_observational_1994}.

The unique chemical signature of \tzos are the key focus of observations aimed at finding these objects. The first TŻO candidate, HV2112, was discovered in the Small Magellanic Cloud \citep{levesque_discovery_2014}. While other candidates have been proposed \citep{beasor_critical_2018,tabernero_nature_2021}, HV2112 has remained the strongest candidate in spite of detailed scrutiny \citep{tout_hv2112_2014,mcmillan_gaia_2018,ogrady_cool_2023}.
 
There are different theories for how these objects can form:  (1) through a common envelope phase where there is double-core evolution \citep{taam_double_1978,ablimit_stellar_2022}, (2) through a stellar collision within a globular cluster \citep{benz_threedimensional_1992}, and (3) a newly formed NS receives a ``kick" that results in a collision with its secondary companion \citep{leonard_new_1994}. Here we refer to the last scenario as the ``impact scenario''. The birth rate for \tzos depends on how they form. The current estimated rate for formation scenario (1) is $ \geq 10^{-4}$ yr $^{-1}$ and for (3) is  $ \sim 10^{-4}$ yr $^{-1}$, leading to an estimated 20 - 200 \tzos in the Galaxy \citep{podsiadlowski_evolution_1995}. While the birth rates for (1) and (3) have been investigated, the estimated birth rate for (2) remains undetermined.

There is ample motivation to search for \tzos and study their interiors and formation. \tzos describe a new fate for massive binary systems which can help us further understand an alternate path of binary evolution. \tzos are a novel channel for producing heavy elements that enrich the interstellar medium due to their unique nucleosynthesis. Studying TŻO progenitors helps us directly understand the relative abundance of these objects, which then further helps us understand which stellar populations and environments would host \tzos. Visually, \tzos are analogous to red giants and supergiants so our search for \tzos directly assists the understanding of giant star physics.

Modeling the formation of \tzos can help us understand the physics of these systems. Since the 2014 candidate discovery, there have been a number of efforts modeling TŻO interiors \citep{farmer_observational_2023} and their formation \citep{hutchinson-smith_rethinking_2023, everson_rethinking_2023}.  While these are important studies into scenario (1), there has been little investigation into the theoretical modeling of TŻO formation through the ``impact scenario''. This study builds on the recent work from \cite{hirai_neutron_2022}, who ran 3D hydrodynamical simulations with varying binary companion masses and kick directions. While they provided a more comprehensive look at newly formed NSs colliding with binary companions, in which they found that certain conditions successfully create \tzos, our study is more explicit in that we specifically examine NS collisions with main sequence (MS) massive stars to determine whether they can successfully create \tzos. By studying the first stage of TŻO evolution, we can gain a better understanding of their population statistics which helps place them in the context of massive star evolution.

In this paper, we investigate the ``impact scenario'' by simulating the mergers of a 1.4 $M_\odot$ NS and a 5 $M_\odot$ to 15 $M_\odot$ MS star.  We assume a MS star because stars spend the majority of their lifetimes in the MS phase (following \cite{leonard_new_1994}). We also assume that the progenitor of the NS was originally the more massive primary, went supernova first, and kicked the NS into an orbit that impacts the MS star. As the only length scale in the problem is the radius of the MS star, $R_{\star}$, we model three periastron distances for the orbit of the freshly kicked NS: 0, 0.5, and 1 $R_{\star}$ . 

We organize the paper as follows. In Section \ref{sec:methods} we outline our underlying assumptions, setup, and initial conditions of our numerical methods. In Section \ref{sec:results} we describe the results of our simulations, concluding with a discussion in Section \ref{sec:discussion}.

\section{Numerical Methods} \label{sec:methods}

\subsection{Setup} \label{sec:setup}

We use the moving-mesh (MM) hydrodynamical solver that has been integrated into the parallel-code Charm N-body GrAvity solver (\changa) \citep{menon_adaptive_2015, wadsley_gasoline2_2017}, named \manga \citep{chang_movingmesh_2017, prust_common_2019}. The algorithm is explained in detail in \cite{chang_movingmesh_2017} and \cite{prust_common_2019}, but we will do a quick overview of it here.

\changa is a parallel N-body smooth particle hydrodynamics (SPH) code that is implemented using the Charm++  parallel language and run-time system.  A recent overview of the SPH algorithm is in \cite{wadsley_gasoline2_2017} and the gravity algorithm is described in \cite{jetley_massively_2008}. The code is unique in that it uses Charm++'s parallelization for communication instead of a typical message passing library. Charm++'s object oriented language has parallel objects executing concurrently with overlapping communication and computation to decompose the gravity and hydrodynamic parts of the calculation. The way \changa takes advantage of the parallelization of the Charm++ language is what sets it apart from other SPH/gravity codes. 

Self gravity is calculated using a tree algorithm that scales as $\mathcal{O}(N \log N)$, where $N$ is the number of particles in the case of SPH or mesh cells in the case of \manga.  The neighbor search needed for the SPH algorithm is also performed using a tree algorithm, again scaling as $\mathcal{O}(N \log N)$.
The gravity algorithm has been shown to scale to 24 billion particles running on 524 thousand cores \citep{menon_adaptive_2015}, and the combined gravity and SPH code has been used to run production cosmology simulations using 2 billion particles on tens of thousands of compute cores \citep{tremmel_romulus_2017}.
The high efficiency minimizes any concerns about performance issues. The tree based gravity and neighbor search algorithms are adaptive in space, and \changa allows each particle to have individual timesteps, making the integration adaptive in time which helps maximize calculation efficiency.

\manga builds on \changa by implementing \cite{springel_pur_2010}'s Arbitrary Lagrangian-Eulerian (ALE) or MM scheme that relies on a Voronoi tessellation. Most MM codes such as TESS \citep{duffell_tess_2011}, ShadowFax \citep{vandenbroucke_moving_2016}, RICH \citep{yalinewich_rich_2015} follow \cite{springel_pur_2010}'s scheme to compute a Voronoi tessellation.  There, they first construct the Delaunay triangulation, and then compute the dual to get the Voronoi tesselation.  \manga takes an alternative approach by directly constructing the Voronoi tessellation by using the Voro++ Voronoi tessellation library \citep{rycroft_voro_2009}. \manga uses the fast neighbor search that is built into \changa to provide neighbors to the Voro++ library.

\manga solves the conservation of mass, conservation of momentum, and conservation of energy Euler equations:
\begin{equation} \label{eq:massE}
\frac{\partial \rho}{\partial t} + \nabla \cdot \rho \mathbf{v} = 0
\end{equation}

\begin{equation} \label{eq:moment}
\frac{\partial \rho \mathbf{v}}{\partial t} + \nabla \cdot \rho \mathbf{vv} + \nabla P = -\rho \nabla \Phi
\end{equation}

\begin{equation} \label{eq:energyE}
\frac{\partial \rho e}{\partial t} + \nabla \cdot (\rho e + P)\mathbf{v} = -\rho \mathbf{v} \cdot \nabla \Phi
\end{equation}
where $\rho $ is the density, $\mathbf{v}$ is the velocity, $\Phi $ is the gravitational potential, $e = \epsilon + \frac{v^2}{2} $ is defined as the specific energy, $\epsilon$ is the specific internal energy, and $P(\rho, \epsilon)$ is the pressure. 

Equations \ref{eq:massE} through \ref{eq:energyE} can be written in a compact form by defining a state vector $\mathbb{U} = (\rho, \rho \mathbf{v}, \rho e)$
\begin{equation} \label{eq:state}
\frac{\partial \mathbb{U}}{\partial t} + \int \nabla \cdot \mathbb{F}dV = \mathbb{S}
\end{equation}
with $\mathbb{F} = (\rho \mathbf{v}, \rho \mathbf{vv}, (\rho e + P)\mathbf{v}) $ being our flux function with V being volume and $\mathbb{S} = (0, - \rho \nabla \Phi, - \rho \mathbf{v} \cdot \nabla \Phi)$ being defined as our source function.

Equation (\ref{eq:state}) is solved using a finite volume scheme as defined by \cite{springel_pur_2010}. \cite{chang_movingmesh_2017} has a further in-depth description of the algorithm, specifically outlining the differences between \cite{springel_pur_2010} and \manga. \cite{prust_common_2019} describes the individual time-stepping algorithm.

\subsection{Initial Conditions} \label{sec:initial}

We create our initial conditions by first constructing a zero-age MS star of a given mass using MESA \citep{paxton_modules_2011, paxton_modules_2013, paxton_modules_2015, paxton_modules_2018, paxton_modules_2019, jermyn_modules_2023}. From this model, we take the resulting mass and radius and construct a polytropic star with polytropic index $\Gamma = 4/3$, matching the mass and radius to within about 1\%. We also assume an ideal gas equation of state in our simulation with no cooling, i.e., $\gamma = 5/3$, where $\gamma$ is the adiabatic index. 

We map the constructed density and temperature profile to a mesh points that will define the simulated star.  These mesh points are constructed by starting out with a cubical uniform mesh.  We then slightly disturb their positions to ensure that the faces are not degenerate and assign masses to each particle such that a SPH smoothing on these particles will produce the appropriate simulated star in \manga.  Note that in \manga as in any other grid code, the region exterior to the star must have some finite density and temperature.  We refer to this region as the ``atmosphere'', and it is given a small density (typically 9-14 orders of magnitude lower than the highest density regions). As this region is large, but uninteresting, we progressively increase the separation between mesh points in this region away from the star.  This reduces the number of mesh points needed to generate the atmosphere, but generally the number of atmospheric mesh points exceeds the number of mesh points used to describe the star.  A detailed description of the algorithm to place these atmospheric mesh points is described in \citet{valsan_envelope_2023}.  

We choose MS masses of $M_{\star} = 5, 7, 9, 12, 15 \ M_{\odot}$. We adopt a minimum mass of 5 $M_{\odot}$ for the MS star. This is in agreement with both the definition of a ``massive" secondary star in a binary system adopted by the BPASS binary stellar population synthesis models \citep{eldridge_binary_2017}, and with the minimum expected mass for a secondary star in a wide binary whose companion was massive enough to die as a supernova progenitor assuming a typical mass ratio (see \cite{moe_mind_2017} and discussions therein). The NS is modeled as a dark matter particle, that is, a point mass that feels only gravitational forces and is not affected by hydrodynamics. We start our simulations with the NS being at a distance of twice the radius of our MS star ($2 \ R_{\star}$). This distance places the NS outside the Roche lobe by a factor of 2 in terms of tidal gravity, which translates to a factor of 8 in gravitational influence, ensuring the initial conditions are not significantly different from an isolated star, but we do not use significant computational resources for the initial approach. We give the NS an appropriate velocity vector to achieve the desired periastron distance, $r_{p}$, set to 0, 0.5, or 1 $R_{\star}$ for a parabolic (energy, $E = 0$) orbit. These values correspond to a direct-collision, envelope-disturbance, and a grazing-encounter respectively.

The choice of an $E = 0$ orbit is a choice of parameter-space expediency. The kick provided by a supernova explosion on the NS will lead to a wide range of orbital energies, including both bound and unbound orbits. Had we chosen either a bound or unbound orbit, we would have been forced to specify a particular total orbital energy or perform a parameter study across different total orbital energies. To limit the parameter space that needs to be explored, we have arbitrarily chosen an $E = 0$ orbit. Future work exploring a possible range of total orbital energies on these NS-MS collisions will need to be informed by population synthesis models to constrain the relevant parameter space.

The gravitational softening length of the NS is 0.1 $R_\odot$, which we determined by a number of numerical experiments to be a good balance between the computational cost of the simulation (since the number of timesteps scales approximately as the inverse of the softening length) and having a small enough value to simulate the behavior of the NS as it plunges into the MS star. Each simulation has approximately 265,000 total mesh points $(N_{\rm tot})$ of which roughly 104,000 mesh points define the initial star, $N_{\star}$.

\section{Results} \label{sec:results}

\begin{figure*}[ht!] 

\center
    
    \includegraphics[width=0.45\textwidth]{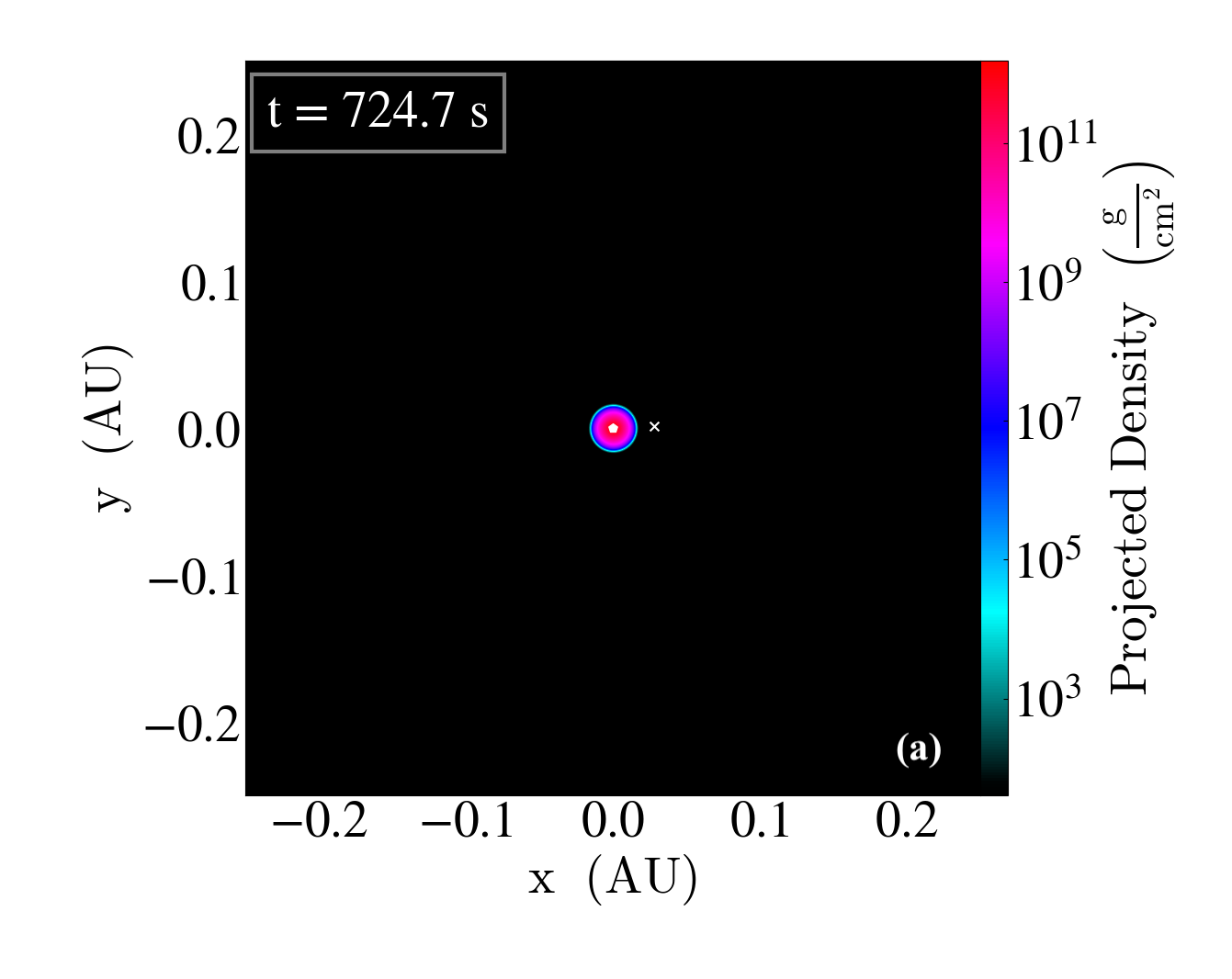} \label{fig:frame1}
    \includegraphics[width=0.45\textwidth]{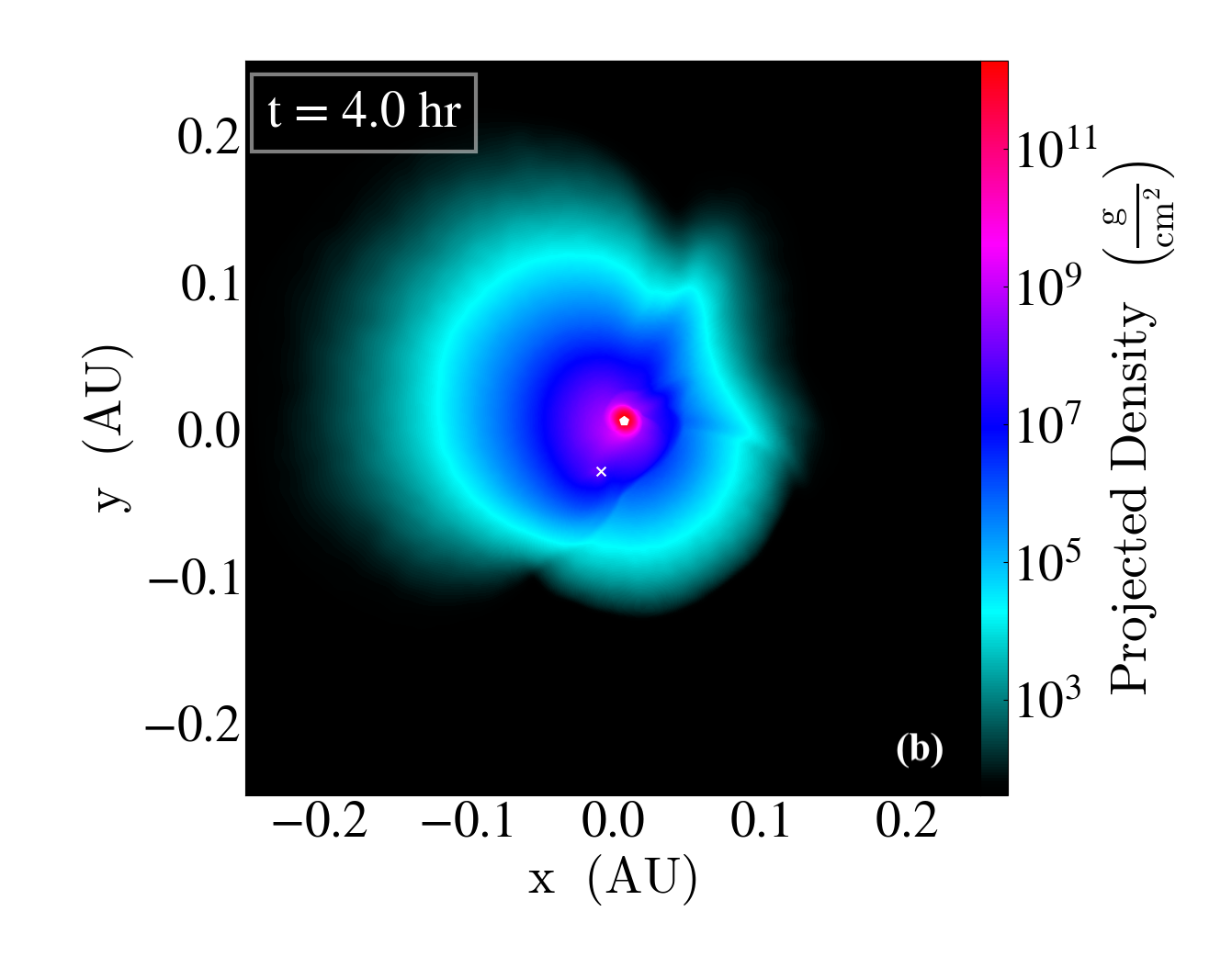} \label{fig:frame2}
    \includegraphics[width=0.45\textwidth]{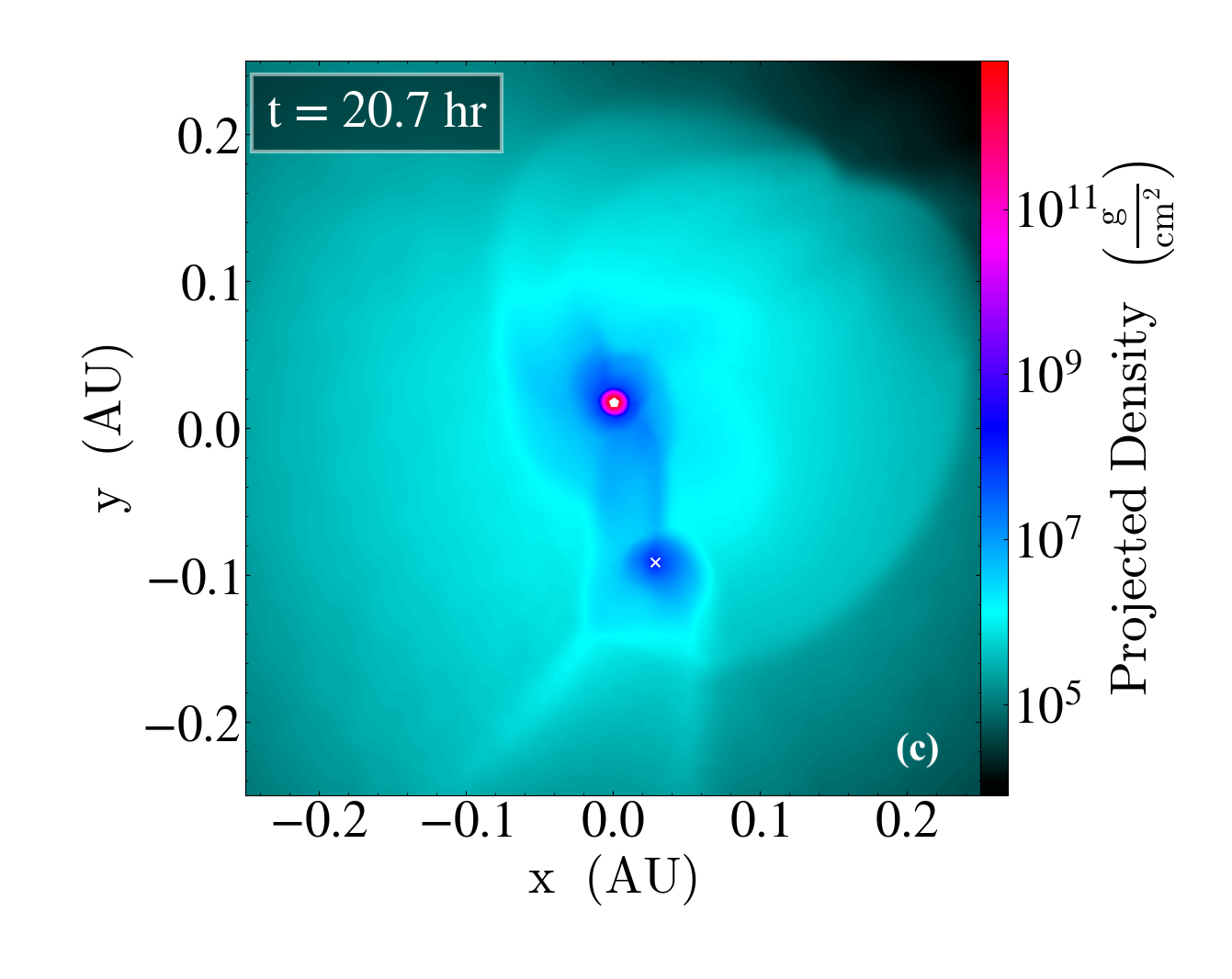} \label{fig:frame3}
    \includegraphics[width=0.45\textwidth]{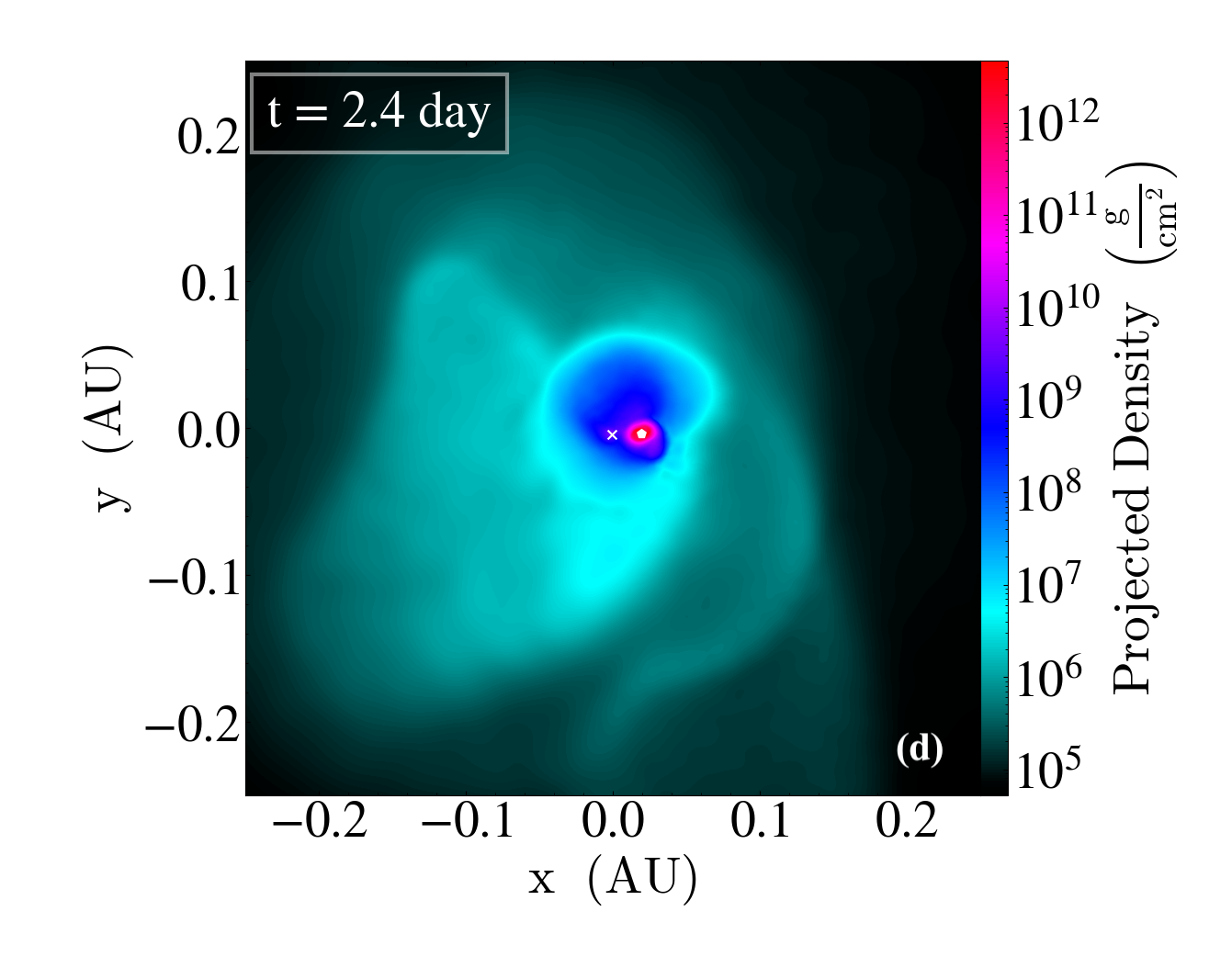} \label{fig:frame4}
    \includegraphics[width=0.45\textwidth]{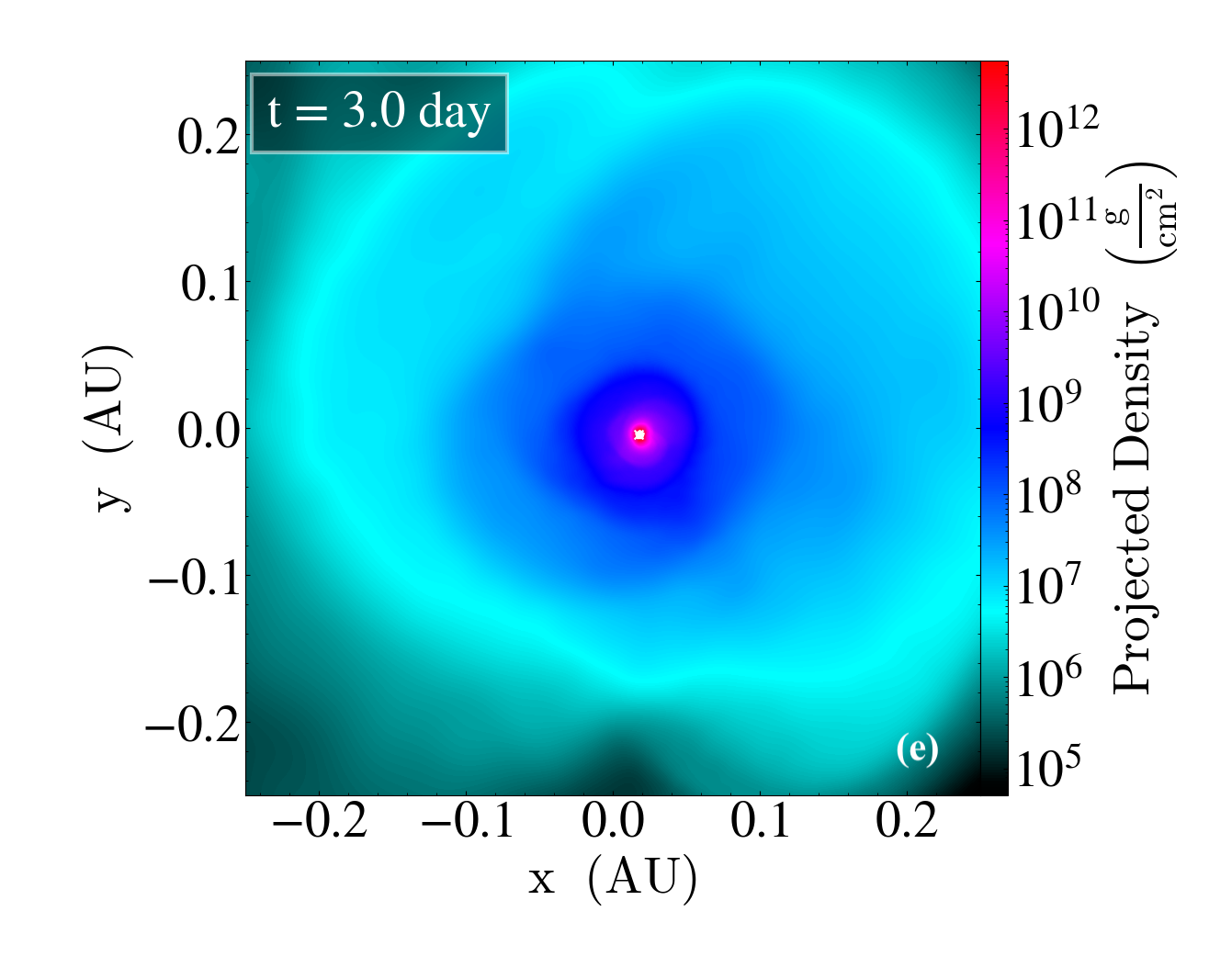} \label{fig:frame5}
    \includegraphics[width=0.45\textwidth]{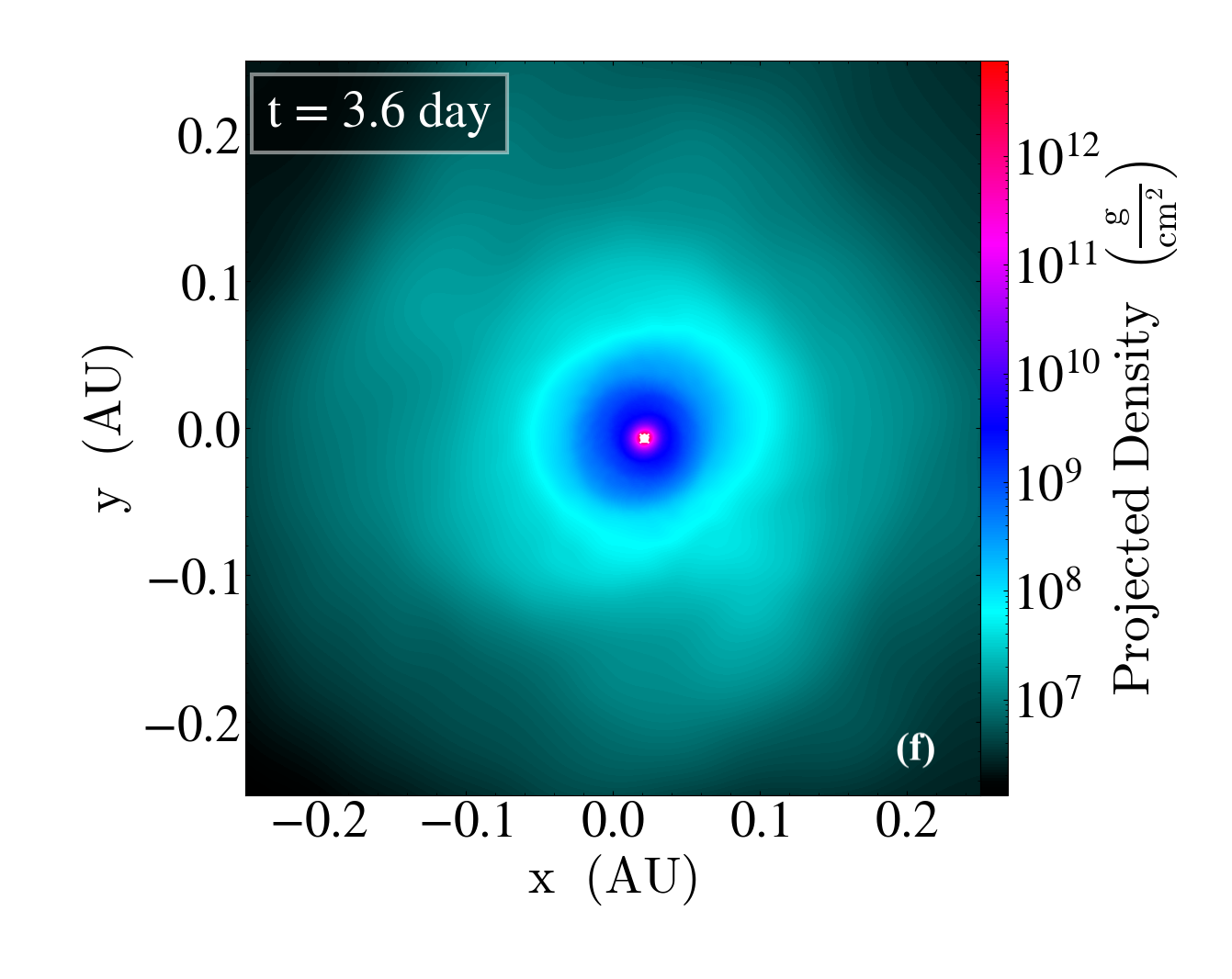} \label{fig:frame6}

    \caption{Frames from our 7 $M_\odot$ MS star merging with the NS: the highest density point, representing the core of the MS star, is indicated by the pentagon and the NS is indicated by the cross. The color bar denotes the density of the gas. The NS successfully merges with the MS star at $t = 3$ days (Frame e).}

    \label{fig:glamour}
\end{figure*}

\begin{figure*}[ht!]

\center
	\includegraphics[width=\textwidth]{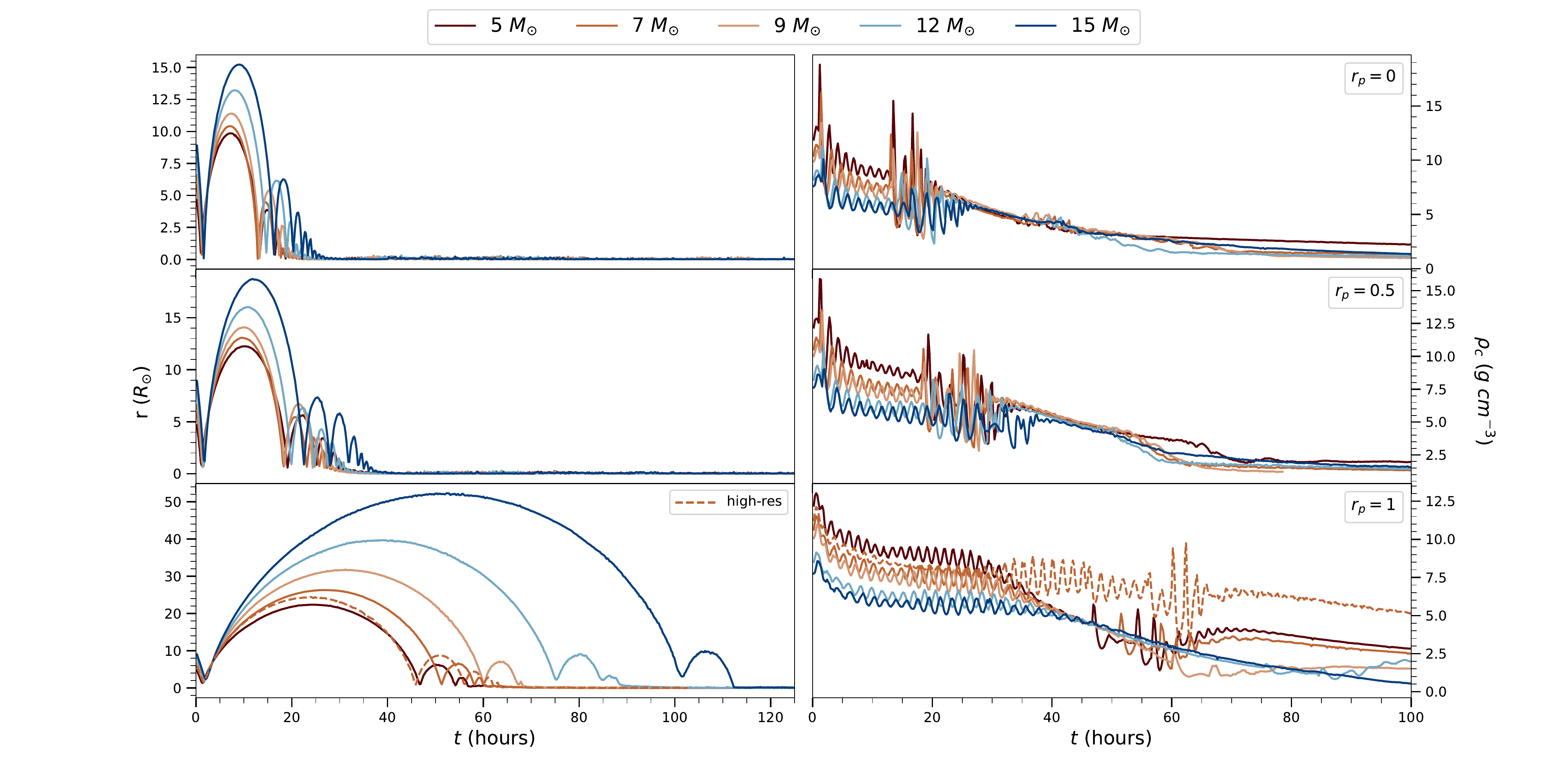}

\caption{Separation between the center of the MS star and the NS, r (in units of $R_{\odot})$, over simulation time is plotted in the left column, with $\rho_{c}$ of our \dtzos plotted in the right column. Each line is colored by the progenitor mass, ranging from 5 $M_{\odot}$ to 15 $M_{\odot}$, and each row corresponds to a different type of encounter: direct-collision ($r_{p} = 0 \ R_{\star}$,top), envelope-disturbance ($r_{p} = 0.5 \ R_{\star}$, middle), and grazing-encounter ($r_{p} = 1 \ R_{\star}$, bottom).}

\label{fig:timeplots}

\end{figure*}

Figure \ref{fig:glamour} shows a series of density projection plots at different times from our high-resolution simulation ($N_{\star}(N_{\rm tot}) = 575 \mathrm{K}(980 \mathrm{K})$) of a 7 $M_\odot$ MS star successfully merging with the NS, and becoming dynamically stable after 3.6 days (35 dynamical times; see Figure \ref{fig:timeplots} for how the system settles into dynamical stability). We define dynamical time, $t_{dyn}$, as the dynamical time for the initial MS star
\begin{equation} \label{eq:dynamical}
    \bm{t_{dyn}} = \sqrt{\frac{R^{3}}{GM}}
\end{equation}
where R is the radius of the star, G is the gravitational constant, and M is the mass of the star. The high resolution simulation was run to obtain a more accurate representation of our merger and to check convergence. Frame 1 (a) starts at time $t$ = 0.2 hours with the separation of twice the star's radius, with $r_p = 1 \ R_{\star}$. Frame 2 (b) shows the initial disturbance of the progenitor at $t$ = 4 hours with the resulting mass loss shown by the density scaling. Frame 3 (c) shows the maximum separation between the NS and the disturbed MS star at $t$ = 20.7 hours with a considerable amount of mass loss surrounding the MS star and the NS. Frame 4 (d) shows the NS at $t$ = 2.4 days continuing to disturb the MS star with each pass as its orbital distance decreases.  Frame 5 (e) shows the successful merger between the NS and the MS star at $t$ = 3 days. We define the two stars to be fully merged when the separation between the MS and the NS is less than the softening length of our simulation, 0.1 $R_{\odot}$. Frame 6 (f) shows the relaxation of the post-merger object at $t$ = 3.6 days, which we define as a dynamical TŻO (dTŻO). Throughout the rest of this paper, we refer to our simulated objects as dTŻOs, this reflects their nature as dynamically formed TŻOs while not yet assuming all the complex physical processes that define a classical TŻO. Our dTŻO is a spherically symmetric, hydrostatic object that maintains its shape and remains dynamically stable for extended periods of time.

\begin{table*}[t!]

	\begin{center}
	\begin{tabular}{
	cccccc}
	
	\toprule

	 $M_{\star, i}$ ($M_{\odot}$) & $r_{p}$ ($R_{\odot}$) &  $t_{m}$ (hrs) & $M_{\star, f}$ ($M_{\odot}$) &  $M_{\star, e}$ ($M_{\odot}$) &  $f_{e}$  \\
	\midrule
	
	\multirow{3}{*}{4.94}  & 0 &19.9& 4.48 & 0.45   & 0.09   \\ 
	
					   & 0.5  &35.2 & 4.62& 0.32 &  0.06\\ 
	
					   & 1  & 69.6 & 4.69 & 0.25 &  0.05  \\

	\hline

	\multirow{4}{*}{6.92} &  0 & 20.3 & 6.43 & 0.49  &  0.07	 \\

					   & 0.5 &30.9  & 6.51 & 0.42  & 0.06  \\ 
					  
					 & 1  &70.8 & 6.62 & 0.30  & 0.04  \\ 
	
					   & 1\textsuperscript{*} &67.3&  6.62 &  0.30 & 0.04 \\

	\hline
	
	\multirow[c]{3}{*}{8.89} & 0 &21.9& 8.44 & 0.45 & 0.05 \\ 
	
					    & 0.5 &31.6 & 8.28& 0.61  & 0.06 \\ 
	
					    & 1 &71.7& 8.66 &  0.23  & 0.03 \\ 
	
	\hline

	\multirow{3}{*}{11.87} &  0 &22.9& 11.54 & 0.32  & 0.03 \\ 
	
					      & 0.5 &32.2 & 11.48 & 0.38 & 0.03 \\ 
	
					     & 1&98.6 & 11.43 & 0.46  & 0.04  \\ 
	
	\hline

	\multirow{3}{*}{14.81} &  0 &26.7 & 14.51 & 0.30  & 0.02 \\ 
	
					      & 0.5 &37.2 & 14.49 & 0.33 & 0.02  \\ 
	
					      & 1 &111.3  & 14.20  &  0.61 & 0.04 \\ 
					    
	\bottomrule
	\end{tabular}
	\end{center}

\caption{$M_{\star, i}$ denotes the initial mass of our recreated star, which is within 1 \% of the corresponding \mesa star. The periastron distance, $r_{p}$, corresponds to different dynamical scenarios explained in Section \ref{sec:initial}. The time to merger, $t_{m}$, is measured from the NS's approach the initial pericenter until the NS and MS star merge; a system is considered merged when their separation falls below the simulation's softening length, 0.1 $R_{\odot}$. The final mass , $M_{\star, f}$, is the mass bound to the \dtzo when $\rho_{c}$ stabilizes. The ejected mass, $M_{\star, e}$ (mass unbound from the \dtzo) is also measured when $\rho_{c}$ stabilizes. The fraction of the initial mass that is unbound is given by $f_{e}$. Our higher resolution run (marked with an asterisk) is numerically converged in both time to merger and bound mass.}

\label{table:1}
\end{table*}

We summarize the results of our simulations in Table \ref{table:1}. The initial mass of the recreated MS star, $M_{\star,i}$, is within 1 \% of the mass of the corresponding \mesa star.  The time from initial pericenter to full merger, $t_{m}$,  is defined when the separation between the MS star and the NS remains at or below the softening length of 0.1 $R_{\odot}$. We define the final mass, $M_{\star,f}$,  as the total mass gravitationally bound to our \dtzo at the time when the central density, $\rho_{c}$, stabilizes. We define the center of our \dtzos as the point of highest density. This is used to determine whether a particle is bound to the \dtzo, when
\begin{equation} \label{eq:unbound}
    \frac{v_p^{2}}{2} + \Phi + \epsilon \bm{\le} 0
\end{equation}
where $v_p$ is the velocity of the particle relative to this center, $\Phi$ is the gravitational potential, and $\epsilon$ is the internal energy. The ejected mass, $M_{\star,e}$, is the difference between $M_{\star,i}$ and $M_{\star,f}$, representing the material that became unbound from the star once the $\rho_{c}$ stabilized. The unbound mass fraction, $f_{e} = \frac{M_{\star,e}}{M_{\star,i}}$, provides a clearer representation of the mass loss, showing that at least 90 \% of the initial mass remains bound to the star. The influence that $r_{p}$ has on $t_{m}$ is evident in the grazing-encounter simulations ($r_{p} = 1 \ R_{\star}$), which take an average of 84 hours to merge, compared with an average of 22 hours for direct-collision simulations ($r_{p} = 0 \ R_{\star}$). The time to merger and final bound mass converge between our high-resolution simulation and the corresponding lower-resolution simulation with the same initial conditions.

Figure \ref{fig:timeplots} shows the separation between the center of the MS stars and the NS over time (left panel), and $\rho_{c}$ of the MS stars (right panel) over time. Identifying the center of the NS is straightforward, as it is represented by a dark matter particle in our simulation. Determining the center of the MS star is more complex. Here we define the center of the MS as the highest-density point within 0.5 $R_{\odot}$ of the center in the previous simulation output.  This approach leads to some uncertainty in the center of the MS stars, as the highest density point is not necessarily  at the star's true center, partly due to numerical noise. Other than being classified by $r_{p}$, there are no other differences in initial conditions between our simulations. All simulations begin at the same relative distance of $2 \ R_{\star}$, and major differences only appear after the NS reaches the initial pericenter (when it first dynamically interacts with the MS star). The MS masses in direct-collision ($r_{p} = 0 \ R_{\star}$) and envelope-disturbance ($r_{p} = 0.5 \ R_{\star}$) simulations reach their maximum separation on similar timescales. In contrast, grazing-encounter ($r_{p} = 1 \ R_{\star}$) simulations take longer to reach their maximum separation. For each $r_{p}$, the higher-mass systems generally require more time to merge than the lower-mass systems, especially in grazing encounters (see Table \ref{table:1}).

We define $\rho_{c}$ as the total gas mass within 0.5 $R_{\odot}$ from the center of the MS star, divided by the enclosed volume. We use $\rho_{c}$ as a proxy for hydrostatic equilibrium. In each simulation, rapid oscillations occur that increase in scale over time, reflecting the dynamic interactions with the NS. Direct-collision and envelope-disturbance simulations evolve on similar timescales. In both cases, oscillations begin and grow in amplitude during the multiple interactions between the NS and the MS star, until the stars merge and then settle for the remainder of the simulation. The grazing-encounter simulations have lower amplitude oscillations and longer durations, consistent with their longer merger times. The flattening of $\rho_{c}$ indicates that the evolution of our \dtzos leads to hydrostatic equilibrium. We note that the high-resolution simulation behaves slightly differently: while $\rho_c$ does not fully flatten (unlike its lower-resolution counterpart), the oscillation duration is similar and is followed by a slow, stable decline, with a factor-of-two difference from the lower-resolution simulation at $t=100$ hours.

\begin{figure*}[ht!]

	\includegraphics[width=\textwidth]{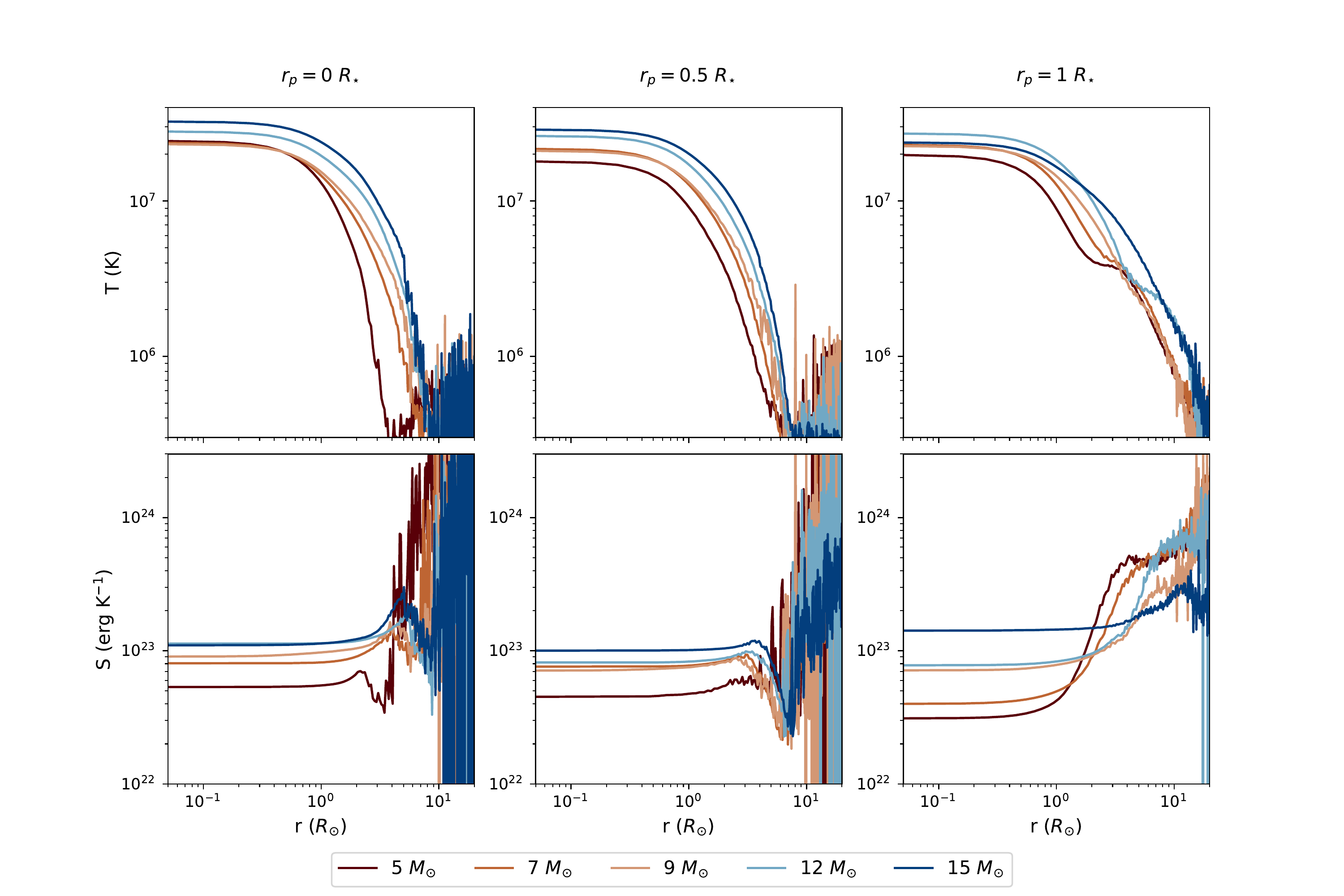}
\caption{Radial temperature and entropy profiles of our \dtzos. The top row shows the temperature profiles, and the bottom row shows the entropy profiles, extending from the center of the \dtzo to approximately $20 \ R_{\odot}$ limited by resolution. The flat entropy profile indicates that the \dtzo is fully mixed or has a convective core. The three columns correspond to one of the three encounter types that we modeled: direct-collision, envelope-disturbance, and grazing-encounter($r_{p} = 0, 0.5, 1 \ R_{\star}$).}
\label{fig:radial}
\end{figure*}

Figure \ref{fig:radial} shows the radial temperature, $T$, and entropy, $s$, profiles of our \dtzos at the time when $\rho_{c}$ stabilizes. The three columns indicate the $r_p$ of the simulations: left, middle, and right columns show $r_p = 0 \ , \ 0.5$, and $1 \ R_{\star}$, respectively. The values of $r_{p}$ are explained in Section \ref{sec:initial}. Our profiles extend to approximately 20 $R_{\odot}$, where the increased noise reflects the limited resolution of the simulations.

The core temperatures of our \dtzos (top row) are well beyond the typical temperatures for CNO burning (though temperatures near the base of the convective envelope in \tzos are expected to be on the order of $\sim10^9$ K, e.g. \citealt{biehle_highmass_1991a, cannon_massive_1993}) and as a result, nuclear reaction and energy generation are crucial in determining the later evolution of these \dtzos. However, modeling the evolution of our \dtzos in thermodynamic and hydrostatic equilibrium with nuclear reactions is beyond the scope of this paper.

Finally, we comment on the entropy profiles (bottom row). 
We define entropy to be 
\begin{equation}
    s = \frac{P}{\rho^{\gamma}}
 \end{equation}
where P is the pressure defined as $\rho\epsilon / (\gamma - 1) $. The entropy profiles are flat to approximately $1 \ R_\odot$ from the center of the \dtzos for all of our simulations. This indicates that our \dtzos are convective or fully mixed in the center, likely due to the NS falling into the core. Convection begins when the entropy gradient is (slightly) negative (the Schwarzchild criterion) and its action drives the entropy gradient to 0.  Conversely, the positive entropy gradient beyond this region implies a stable outer envelope. Compared direct-collision and envelope disturbance simulations, the grazing-encounter simulations exhibit steeper entropy gradients. This likely reflects less efficient mixing in the interior of the envelope, due to the higher number of NS interactions over the the longer merger times.


\section{Discussion \& Conclusion} \label{sec:discussion}

In this paper, we used \manga, a moving-mesh hydrodynamical code integrated into \changa to demonstrate that \tzos can robustly form from a NS-MS merger for small periastron distances, $r_{p} \le 1 \ R_{\star}$. The merger will result in a merged object that is stable against dynamical collapse, which we define as a \dtzo.

We modeled three periastron distances to explore a range of possible dynamical interactions: direct-collisions ($r_{p} = 0 \ R_{\star}$), envelope-disturbances ($r_{p} = 0.5 \ R_{\star}$), and grazing-encounters ($r_{p} = 1 \ R_{\star}$). Our simulations show that the increased $t_{m}$ for grazing encounters can influence the settling behavior of the \dtzo, as evidenced by the range and timescales of oscillations in $\rho_{c}$ (see Figure \ref{fig:timeplots}). The internal temperatures of our \dtzos are initially very high, exceeding typical CNO burning temperatures. This implies that the physics of radiation and nuclear burning must be taken into account. The entropy profiles indicate that the interiors of our objects are likely convective or fully mixed, with a stable outer envelope. While our models do not yet include the full range of physical processes specific to TŻOs, this work represents an important first step using a powerful and flexible code capable of capturing the key dynamical features.

Future research will include simulating a range of periastron distances larger than $1 \ R_{\star}$ to further examine how robustly \tzos can form from a variety of NS-MS collisions, and to examine other products of these binary interactions \citep{hirai_neutron_2022}. By expanding the range of periastron distances, we can gain a clearer understanding of how initial conditions influence the interactions and dynamics of these systems. Our simulations can further be utilized as initial conditions for stellar evolution codes to see the long-term evolution of these objects.

Our models demonstrate that \tzos can indeed be dynamically formed in the ``impact scenario''. When integrated in binary population synthesis calculations, our results can help estimate the population of \tzos formed through this pathway. This study further suggests that the ``impact scenario'' may represent an important alternative terminal pathway for massive binaries that might otherwise be presumed as NS-NS gravitational wave progenitors.

\begin{acknowledgments}
We thank the anonymous referee for their valuable comments that helped improve the manuscript. LEW is supported by NSF-GRFP grant DGE-214004 and NSF grant AST-2305224 awarded to EML. PC acknowledges support from the NSF via grants AST-2307885 and AST-2108269 and the NASA ATP program (NNH23ZDA001N-ATP). Computations were done on the mortimer supercomputer at UWM, which is funded by UWM and the NSF through grant OAC-2126229. We use the yt software platform for the analysis of the data and generation of plots in this work \citep{turk_yt_2011}. This research has made use of NASA’s Astrophysics Data System. The Scientific color map vik \citep{crameri_scientific_2023} is used in this study to prevent visual distortion of the data and exclusion of readers with color-vision deficiencies \citep{crameri_misuse_2020}. 
\end{acknowledgments}


\clearpage

\bibliographystyle{aasjournalv7}
\bibliography{Williams_arxiv}

\end{document}